\documentclass[preprint,12pt]{aastex}

\slugcomment{Accepted for publication at The Astrophysical Journal}

\shorttitle{Monolithic Emission in Barnard 59}
\shortauthors{Rom\'an-Z\'u\~niga et al.}

\begin{document}


\title{Barnard 59: No Evidence for Further Fragmentation}


\author{C. G. Rom\'an-Z\'u\~niga}
\affil{Instituto de Astronom\'ia, Unidad Acad\'emica de Ensenada, Universidad Nacional Aut\'onoma de M\'exico, Km 103 Carr. Tijuana-Ensenada, Ensenada BC 22860 MEXICO}
\email{croman@astrosen.unam.mx}

\author{P. Frau and J. M. Girart}
\affil{Institut de Ci\`encies de l$'$Espai (CSIC-IEEC), Campus UAB, Facultat de Ci\`encies, Torre C-5p, 08193 Bellaterra, Catalunya, Spain}

\and

\author{Jo\~ao F. Alves}
\affil{Institute of Astronomy, University of Vienna, T\"urkenschanzstr. 17, 1180 Vienna, Austria}




\begin{abstract}
The dense molecular clump at the center of the Barnard 59 (B59) complex
is the only region in the Pipe Nebula that has formed a small,stellar cluster. The previous analysis of a high resolution near-IR dust extinction map
revealed that the nuclear region in B59 is a massive, mostly quiescent clump of 18.9 M$_\odot$. The clump shows a monolithic profile, possibly indicating that the clump is on the way to collapse, with no evident fragmentation that could lead to another group of star systems. In this paper we present new analysis that compares
the dust extinction map with a new dust emission radio-continuum map of higher
spatial resolution. We confirm that the clump does not show any significant evidence for prestellar fragmentation at scales smaller than those probed previously. 

\end{abstract}


\keywords{stars: formation --- ISM: clouds --- radio continuum: ISM --- infrared: ISM --- radio lines: ISM}



\section{Introduction \label{s:intro}}

\par The formation of a stellar cluster proceeds when a molecular cloud clump hosts multiple dense fragments capable of collapsing independently. Cluster-forming clumps populate the high-end bins of the mass and density distributions in a cloud \citep{Williams:1995aa,Di-Francesco:2010aa}, but very few details are known about the process of fragmentation, since cluster-forming clumps in very early stages of evolution are relatively rare.

\par The recently popular Pipe Nebula \citep[see][and references therein]{Alves:2008kx}, has become a prototype case for a cloud in a very early stage of evolution. The Pipe Nebula hosts only one star cluster-forming clump, Barnard 59 (B59), and one core hosting a single young source in a nearby filament \citep{Forbrich:2009ab}. The rest of the cloud contains more than 130 starless cores that resemble stars in the way their masses are distributed \citep{Alves:2007aa,Rathborne:2009aa} and how they are distributed spatially \citep{Roman-Zuniga:2010vn}. Dense cores in the Pipe appear to be still mostly quiescent and stable against collapse \citep{Lada:2008aa}, despite some of them being already chemically evolved \citep{Frau:2010fk,frau11a,frau11b}. Currently, the best candidate for a mechanism that supports cores in the Pipe against collapse is the magnetic field that appears to permeate the cloud \citep{Alves:2008bk,Franco:2010ly}. 

\par The B59 complex hosts one of the less massive and less distant ($d=130$ pc) young stellar clusters observable. During the last 2.6 Myr, B59 has formed 14 stars, all below 3 M$_\odot$ \citep[][hereafter CLR10]{Covey:2010uq}. The analysis of a high-resolution (24$\arcsec$) near-infrared dust extinction map of the B59 region \citep[][hereafter RLA09]{Roman-Zuniga:2009aa} revealed that B59
is a complex group of dense cores and filamentary structures, in which
the central clump, B59-09, hosts most of the cluster members. The analysis
of the dust extinction map suggests that the central clump has a smooth profile that compares well with that of an isothermal sphere, with no evidence of internal fragmentation. Moreover, pointed NH$_3$ observations inside the core show that the thermal to non-thermal kinetic energy ratio averages well over unity, suggesting that B59-09 remains mostly quiescent despite having formed a star cluster. The main goal of this study is to confirm such a scenario, by showing that a radio continuum dust emission map with higher spatial resolution, also lacks evidence for fragmentation in the central core. 

\par The paper is divided as follows: in $\S$\ref{s:obs} we describe the observations and data reduction process. Our results are detailed in $\S$\ref{s:analysis} and, to conclude, we elaborate a discussion about their significance in $\S$\ref{s:discussion} .

\section{Observations \label{s:obs}}

\subsection{MAMBO-II \label{s:obs:ss:mambo}}

We mapped B59 at 1.2 mm (250 GHz) with the MAMBO-II bolometer at the 30m IRAM telescope atop Sierra Nevada (Spain). MAMBO-II features a 117-receiver array that covers 240$\arcsec$ in diameter. The observations were carried out in November 2009 in the framework of a flexible observing pool. The weather conditions were good and zenith optical depth remained within 0.1 to 0.3. A total of 5 usable on-the-fly maps were completed and combined. The beam size of the telescope is 11$\arcsec$ at the effective frequency of 250 GHz; we used a constant scanning at a speed of 8$\arcsec\ \mathrm{s}^{-1}$ in the azimuthal direction for up to 65s. This strategy resulted in average integration times per map of $\sim$1 hour. Each map was performed with a different secondary chopping, which varied between 30$\arcsec$ and 80$\arcsec$, parallel to the scanning direction of the telescope. This procedure assured that we had a different OFF position for each ON position within the map. The scanning direction was also changed for each map, giving a different zero emission level, which helped to 
avoid spatial filtering effects. The zenith optical depth was measured with a skydip at the start and end of a map. Pointing and focus were checked also at the start and end of a map, with corrections below 3$\arcsec$ and 0.2 mm, respectively. Flux density calibrators were observed every few hours. The data reduction was done with standardized routines from MOPSIC software included in the GILDAS package\footnote{GILDAS and MOPSIC are available at http://www.iram.fr/IRAMFR/GILDAS}.

\subsection{Near Infrared Dust Extinction Map and additional radio data \label{s:obs:ss:nicer}}

\par We make use of the dust extinction map of RLA09 constructed with the NICER technique \citep{Lombardi:2001aa} and a combined photometric catalog obtained from ground (ESO) and space based (Spitzer) observations. In addition to the infrared data, we also make use of a series of pointed observations made with the Green Bank 100m telescope to determine the variation of the emission of the (1,1) and (2,2) rotation-inversion transitions of ammonia (NH$_3$) across the central clump B59-09. These data were also used and described in RLA09. Finally, we also make use of a C$^{18}$O (2-1) line emission map obtained with the detector HERA at the IRAM 30m telescope (Rom\'an-Z\'u\~niga et al., in prep). 

\section{Data Analysis and Results \label{s:analysis}}

\par
\begin{figure}
\includegraphics[scale=.5]{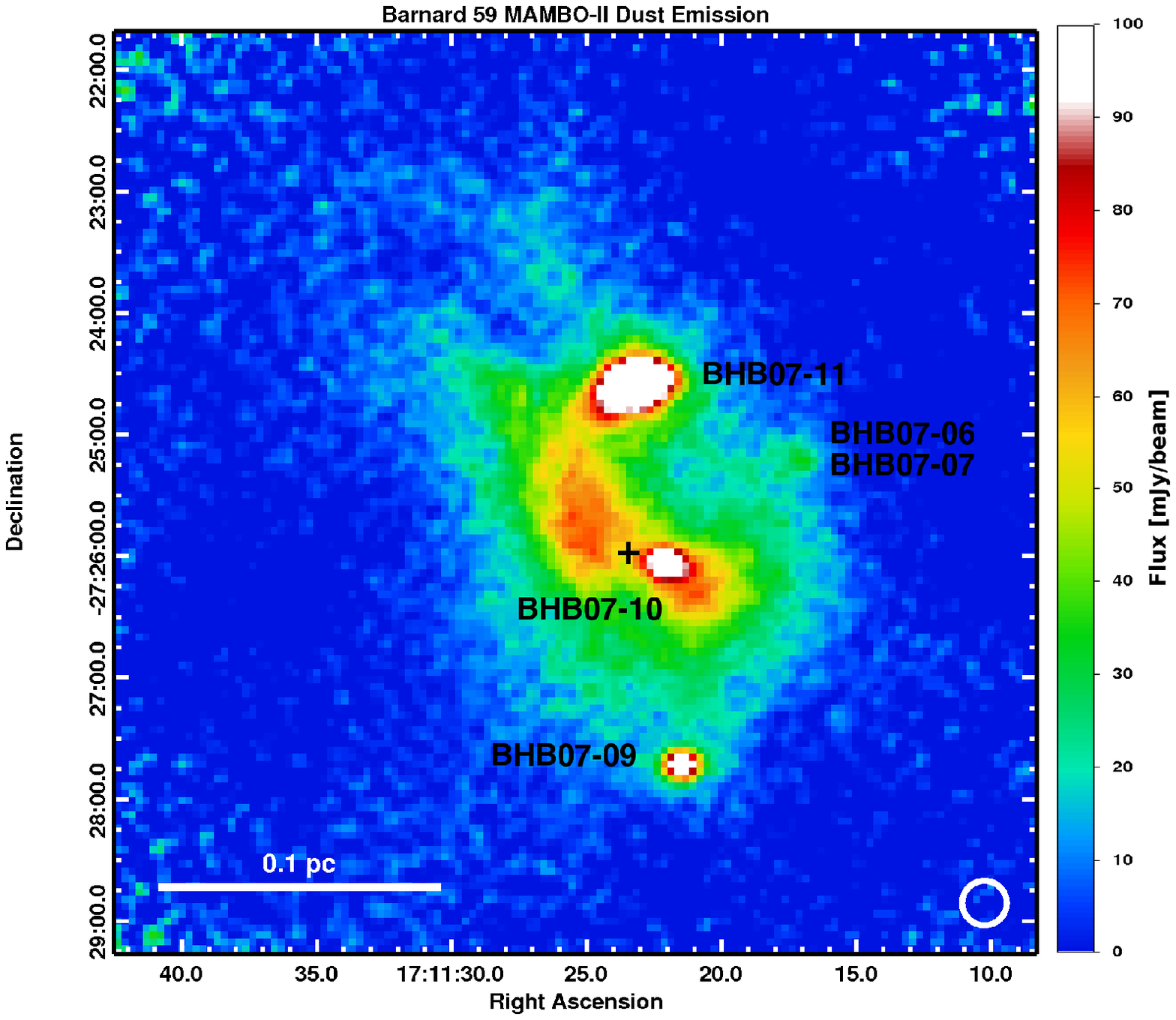}\hspace{0.1in}\includegraphics[scale=.5]{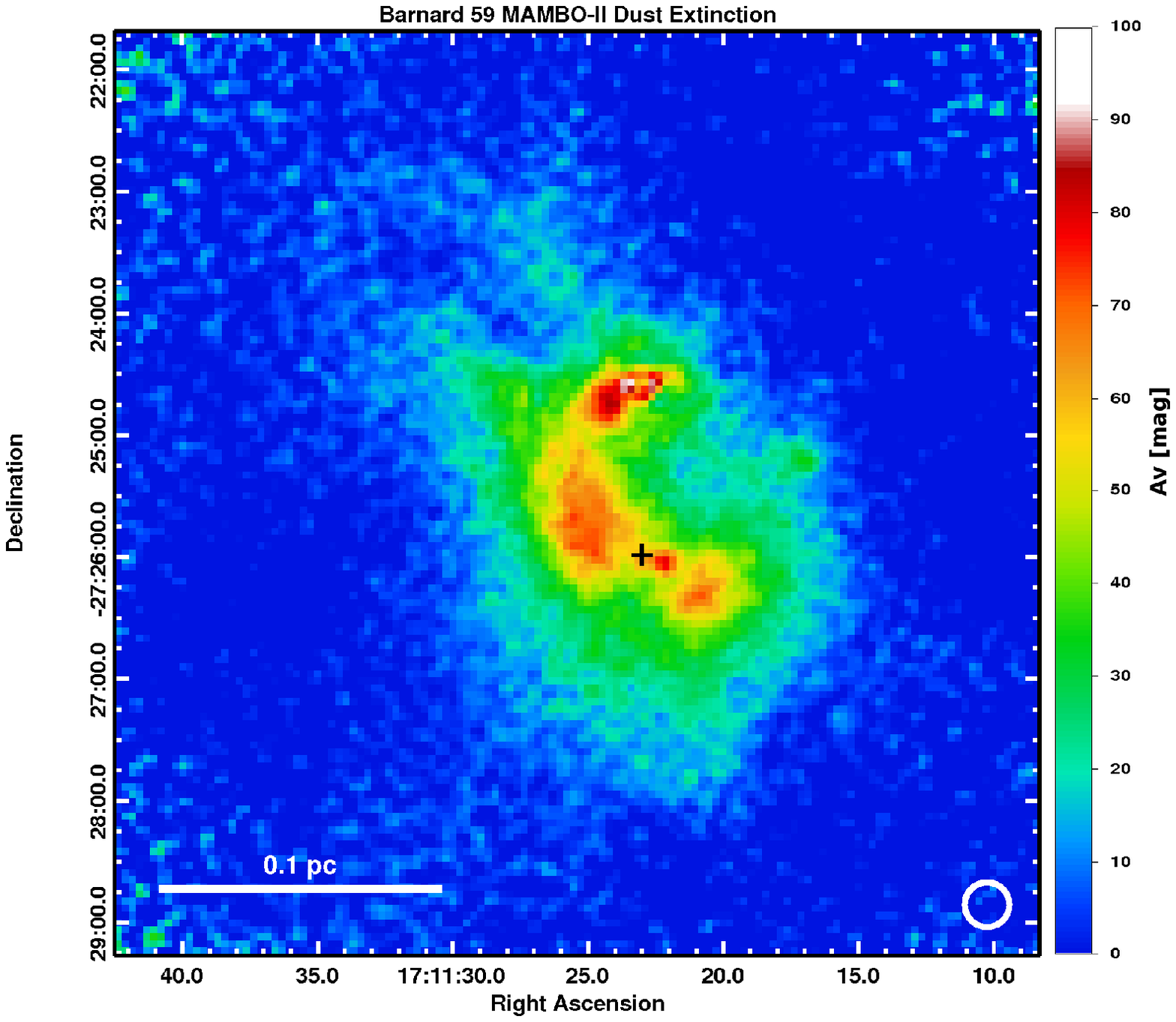}\\
\includegraphics[scale=.5]{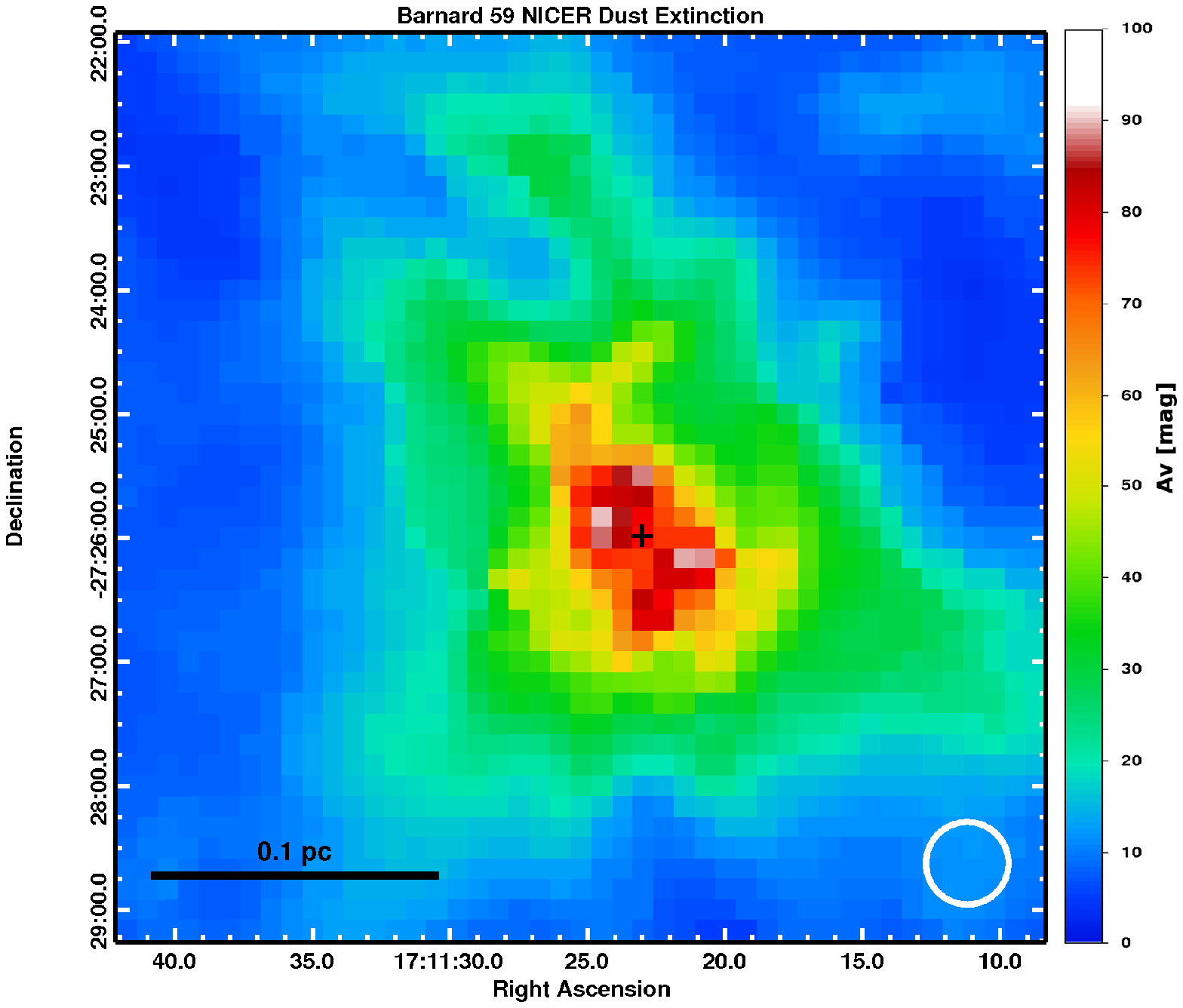}
\caption{\textit{Top Left:} MAMBO-II dust emission map of Barnard 59 from our observations. For purposes of clarity, the flux scale has been clipped from 0 mJy/beam to 100 mJy/beam. Numbers refer to YSO sources identified and listed in Table \ref{tab:ysos}. The circle at the bottom right corner indicates the beam size. \textit{Top Right:} MAMBO II Dust extinction map after conversion of the flux and subtraction of the YSOs emission. The extinction level has been also clipped from 0 mag to 100 mag. The circle at the bottom right corner indicates the beam size. \textit{Bottom:} NICER dust extinction map, as in RLA09, also scaled from 0 mag to 100 mag. The circle at the bottom right corner shows the size of the Gaussian filter used to construct the map. In all panels, the cross symbol marks the center around which radial profiles were constructed, at J2000 $(\alpha,\delta)=\mathrm{(17:11:23,0,\ -27:25:59.3)}$. \label{fig:maps}}
\end{figure}

 In the top panel of Figure \ref{fig:maps} we show our MAMBO-II map toward B59. The map detected mostly the emission from the clump 09ab\footnote{we follow the nomenclature of RLA09, see their Table 1} and, less prominently, the core 09c, at the northwestern end. Five young stellar objects were detected with very high signal-to-noise ratios (SNRs). They correspond to sources 6, 7, 9, 10 and 11 in the list of \cite{Brooke:2007aa} (BHB07). The 250 GHz continuum emission properties of these sources are listed in Table \ref{tab:ysos}. Notice that sources BHB07-6 and BHB07-7 lie very close together and they are not resolved as separated sources in the MAMBO-II map. As our main purpose is to study the emission of the core, we subtracted out the contribution of the young stars. For this purpose, a 2D Gaussian profile was fit to each of the sources, using a background emission level corresponding to the average in the vicinity region of the clump. The fit parameters are listed in Table \ref{tab:ysos}. The smoothness of the resulting map (see central panel in Figure \ref{fig:maps}) seems to confirm that most of the subtracted emission arises from the YSO warm circumstellar material rather than from the core cold dust.

\par After subtracting the contribution from the YSOs, we transformed the dust emission maps to column density and then into visual extinction for a direct comparison with the NICER map of RLA09. The conversion was done following the method described in section 4.1 and appendix A of \cite{Frau:2010fk}. We notice that the conversion is very sensitive to the assumed value for gas temperature, $T_k$ \citep[see Appendix 1 of][]{Frau:2010fk}, and also the dust emissivity ($\kappa_\nu$) and the dust temperature assumed. Following \citet{ossen94}, we assumed $\kappa_\nu$=0.007~cm$^2$~g$^{-1}$ for grains in a dense medium ($n$$\sim$ 10$^5$~cm$^{-3}$). Then, we computed visual extinction maps for gas/dust temperatures in the 10--12~K range in steps of 0.25~K \citep[ammonia observations yield $T_K$=11.3 K$\pm$0.7 K;][]{Rathborne:2009aa}. We selected a temperature of 10.25~K as this value showed the best agreement with respect to the previous extinction map, and we made the assumption that this temperature is constant within the entire clump. The derived radius, mass and average density of the clump are listed in Table \ref{tab:clump}. Given the high SNR achieved, we restricted analysis to the region satisfying $I_\nu$$>$$0.2~I_{\rm \nu, max}$ ($\sim$4.5$\sigma$). The total mass of the clump 09ab in B59 estimated from the MAMBO-II map is about half of that estimated from the NICER map (i.e., cores 09ab and 09c sum about 21 M$_\odot$ according to RLA09). This difference is mostly because MAMBO-II maps are not as sensitive as the near-infrared excess method in detecting the diffuse gas at the external parts of dense cores ($A_V$$<$25~mag in this case), rather than a technical effect. The partial detection of cores 09cd and 04a seem to confirm the good quality of the emission maps despite their lack of sensitivity. The NICER technique relies on having enough sources per beam to average reddening, and background sources toward the edges of B59 are abundant. On the other hand, the continuum emission relies on the amount of dust that contributes to the signal, which decreases at lower column densities \citep[see][for additional discussions on this effect]{Frau:2010fk}. Also, the assumption of a constant temperature for the entire region is less accurate toward the core boundaries that might be heated by external sources.

\par
\begin{figure}
\plotone{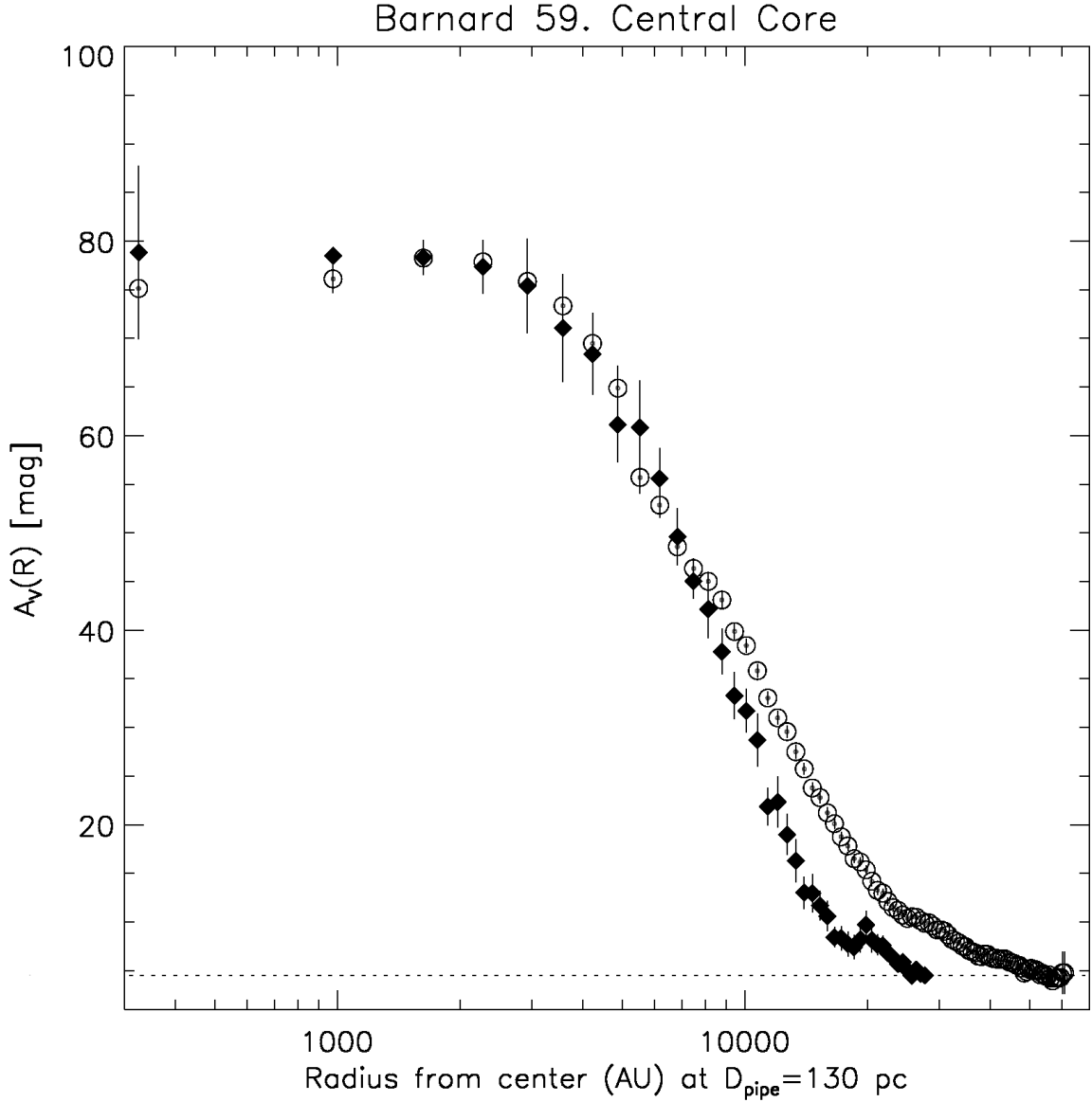}
\caption{The open circle symbols show the radial profile of the NICER extinction map. The black diamond symbols show the radial profile of the MAMBO-II dust emission map. The profiles were constructed by averaging flux in circular, concentric rings centered on J2000 $(\alpha,\delta)=\mathrm{(17:11:23,0,\ -27:25:59.3)}$. For the MAMBO-II profile, the data were convolved to the resolution of the NICER map (20$\arcsec$) and re-gridded to match pixels in both maps. \label{fig:profile}}
\end{figure}

 Morphologically, the MAMBO-II dust extinction map shows features that are equivalent features to those found in the NICER map of RLA09, which is shown 
in the bottom panel of Fig \ref{fig:maps}. Both maps show a relatively flat central region toward the clump 09ab with a shallow ``dent" near the location of source BHB07-10, discussed in section 5.1 of RLA09. A radial profile of the MAMBO-II map was constructed by averaging flux in circular, concentric rings centered on J2000 $(\alpha,\delta)=\mathrm{(17:11:23,0,\ -27:25:59.3)}$, as in RLA09.  Figure \ref{fig:profile} shows that observed profiles compare well to each other and trace equivalent structures within $\sim$10$^4$~AU ($A_V$$>$35~mag). Below $A_V\approx35$ mag, the MAMBO profile plunges and appears to be truncated below an average radius of 2$\times10^4$ AU, while the NICER profile continues until it reaches the background level near 5$\times10^4$ AU. 

\par We used our pointed NH$_3$ observations (RLA09) and the C$^{18}$O(2-1) map to categorize the variation of the LSR velocity near the center of the clump. Twelve NH$_3$ pointed observations were made within the central 10$^4$~AU of B59. All of them indicate variations of $v_{LSR}$ smaller than 0.088 km s$^{-1}$ from the central value of 3.485 km s$^{-1}$. Moreover, these variations are below the average velocity dispersion $\langle\sigma_v\rangle=0.210\pm0.0498$  km s$^{-1}$ and below the sound speed in a 10 K gas ($c_s=0.12\mathrm{\ km\  s}^{-1}$). The C$^{18}$O (2--1) map reveals variations below 0.05 km s$^{-1}$ from the central velocity value. The linewidths, although being significantly wider than those of the NH$_3$, also show very small variations (less than 0.08 km s$^{-1}$) from the average value of 0.87 km s$^{-1}$, indicating very uniform kinematics near the center of B59-09ab. \cite{Rathborne:2009aa}, discussed how kinematically independent cores in  the Pipe Nebula must show radial velocity differences larger than $c_s$. Following this prescription, the differences we observe are too small to suggest the presence of kinematically independent substructure. 

\section{Discussion \label{s:discussion}}

\par The radio continuum continuum map has twice the resolution of the NICER map (11$\arcsec$ FWHM beam vs. 24$\arcsec$ FWHM Gaussian filter), resolving a projected size of 1420 AU (assuming a distance of 130 pc to the Pipe Nebula). We could safely say that the MAMBO-II map should resolve structures with minimum sizes of 2000 AU to 5000 AU. Our map, however, does not reveal any significant substructure (other than the sources or the dent) below $\sim1.5\times\mathrm{10}^4$ AU. This result is in good agreement with those of \citet{Roman-Zuniga:2010vn}, who
reported a possible limiting scale of fragmentation in the Pipe Nebula of
about $1.4\times\mathrm{10}^4$ AU. As also noted there, this scale agrees with the results of \citet{Schnee:2010uq} who found no evidence of fragmentation in cores of the Perseus Molecular Cloud at scales of $10^3-10^4$ AU.

\par The column density estimations made from near-IR extinction and dust emission are very similar near the center. Both maps fail to show evidence of substructure suggestive of significant fragmentation in R09ab. The dust emission map confirms that the dent is real and thus it suggests that source BHB07-10 is possibly affecting the material surrounding it. The decrement of column density near source BHB07-10, however, is equivalent to a decrement of only 3 percent in the total mass of the clump (RLA09). Also, CLR10 showed that BHB07-10 is not embedded at the level at which it is projected against the map. In contrast, source BHB07-11 has been suggested to be the origin of a moderate gas outflow \citep{Onishi:1999aa} that seems to be carving the northern part of the clump, forming core 09c. Thus, the feedback effect near BHB07-10 may not account for a bonafide fragmentation process as in core 09c. 

The young stellar cluster in B59 is likely too small to hold itself together dynamically. Following \citet{Adams:2001aa}, for the B59 cluster to remain stable, its relaxation timescale, $t_{rlx}$, should be at least larger than its formation timescale, which in turn should be comparable to the age of the cluster. The age of the B59 cluster is estimated as being 2-3 Myr (CLR10). The relaxation parameter, $Q_{rlx}$, i.e., the number of crossings per relaxation time, can be estimated in terms of the star forming efficiency, $\epsilon$, and the number of stars, $N_\star$, approximately as $N_\star/(10\epsilon^2\ln{(N_\star/\epsilon)})$. Using the present day value, $\epsilon=0.3$ (CLR10) and, for 10 sources projected within B59-09ab, 
we obtain $Q_{rlx}=3.2$. Thus, given the crossing time, of the clump $t_{cr}\approx0.06$ Myr, and $t_{rlx}=Q_{rlx}t_{cr}\approx 0.2$ Myr, which is much shorter than its formation time. It is thus very unlikely that the B59 stellar cluster will remain together longer than a few more crossing times. 

\par Experimental fits of isothermal sphere profiles, particularly Bonnor-Ebert and Dapp-Basu \citeyear{dapp:2009aa} to the NICER profile suggest that the core is already out of equilibrium, although it is not possible to estimate its state of evolution toward collapse (see also RLA09). The velocity dispersion from the NH$_3$ data and the parameters from Table \ref{tab:clump} yield a virial parameter value $\alpha_{vir}=0.25$, which suggests that the B59-09ab clump is gravitationally bound. Thus, we should expect the clump to be presently moving towards collapse. As shown in RLA09, B59-09ab remains mostly quiescent. Moreover, our data show that the clump has been able to amass between 9 M$_\odot$ and 19 M$_\odot$ of gas (the latter if we consider the whole dust extinction structure) in an apparently monolithic structure, and we know from the age of the cluster that the period of mass aggregation could be as long as about 6 crossing times long (CRL10). What is the mechanism that held the clump together, retarding collapse and fragmentation for a relatively long period?

\par Feedback in the form of outflows could serve as a turbulent
energy injection mechanism that could provide the non-thermal support required
to maintain the clump against collapse. Several authors have
provided evidence of one and possibly two outflows from embedded sources BHB07-09 and BHB07-11 \citep{Onishi:1999aa,Brooke:2007aa,Riaz:2009dy}, and RLA09 discussed how these could be carving structures at the outer regions of the clump. The presence of at least 4 YSOs located at projected distances larger than the Jeans length of the core (see below) could suggest that B59-09ab is more likely a remnant of dense gas after an episode of multiple source formation within a much larger structure. Even source BHB07-10 could also be affecting the core at a shallow level, as discussed above. The numerical experiments of \citet{Krumholz:2007hs} and \citet{Offner:2010ks} have shown that radiative feedback can be a strong agent against fragmentation. For instance, feedback from protostars can inhibit the formation of binaries in a fragmenting disc scenario. These models, however, require stars with masses above 3 $M_\odot$ and the mechanisms work best at spatial scales about one order of magnitude smaller than the observed size of B59. Since sources in B59 have too low masses, and are not numerous, it is thus very unlikely that they can either or provide non-thermal support against collapse neither provide enough feedback energy to inhibit fragmentation \citep[see, e.g.,][]{Longmore:2010ei}.

\par The other plausible candidate for a mechanism that can prevent further collapse and fragmentation of cores in B59 (as well as other regions of the Pipe Nebula) is the magnetic field. Numerical studies like those of
\cite{Nakamura:2011hs} concluded that for an initially magnetically-subcritical
cloud, a strong magnetic field is able to slow down gravitational collapse
and fragmentation, decreasing the star formation rate significantly. Also, \cite{Price:2007el} showed that support by magnetic fields may deter density perturbations and the fragmentation of discs in the case of binary formation. The optical polarization studies of \cite{Alves:2008bk} and \cite{Franco:2010ly} strongly suggest that the Pipe is permeated by a magnetic field. Moreover, \citet{Frau:2010fk} have shown that chemically evolved cores in the Pipe are possibly associated with a strong magnetic field, which is suggestive of significant magnetic support. 

\par Following the formulation of \citet{Mouschovias:1991aa}, we find a Jeans length scale of $\lambda_{T, cr} = 0.12$~pc, which is about twice as large as the radius of the clump in the dust emission map, within which have demonstrated the monolithic behavior of the clump (although this length is very close to the radius estimated from the dust extinction map). The data of \citet{Franco:2010ly} indicates that B59 may be sub-Alfv\'enic, so in the overall region the magnetic field is dynamically more important than the turbulence.

From the Jeans length scale and the critical magnetic length scale, $\lambda_M=0.91(B/\mathrm{\ }[\mu\mathrm{G}])(1\times 10^3\mathrm{\ [cm}^{-3}\mathrm{]}/n)$ \citep{Mouschovias:1991aa}, and considering both the visual extinction and dust emission map derived values of the clump radius, we can infer that a magnetic field strength in the range of 0.1--0.2~mG would be enough to support the clump. These values are larger than those calculated from optical polarization data of the diffuse surrounding gas. However, magnetic field is expected to strengthen toward denser regions as the collapse process evolves \citep[e.g.][]{Fiedler:1993fk} and these estimates are not unreasonable compared to other dense cores \citep[see][]{Crutcher:1999aa,Crutcher:2004aa}.

While the profile of B59 suggests the core is out of equilibrium (Rom\'an-Z\'u\~niga et al., in prep), our continuum map shows that it has not finished collapse for a time comparable to the age of the stellar cluster. Magnetic field support (or an equivalent combination of supporting mechanisms) could have been active during such a timescale. The age of the cluster suggests that B59-09ab has 
survived for a period longer than 10 $t_{ff}$ without completely collapsing or 
fragmenting, and it is unlikely that it will fail to evolve toward protostellar
collapse. Experiments by \citet{GalvanMadrid:2007vy} suggest that pre-stellar core survival can be assured for at most 3-10 $t_{ff}$ for cores with densities above $10^5\mathrm{\ cm}^{-3}$, independently of mass to magnetic flux ratio. Despite the lack of fragmentation in B59-09ab at present, we cannot assure from our data only that the clump will not fragment later. We think, however, that further fragmentation is unlikely because fragmentation tends to proceed quite rapidly. For instance, some numerical studies, like those of \citet{Boss:2009vn}, show that oblate cores with magnetic fields can form binaries in timescales of less than 2-4 $t_{ff}$. Also, the  models of \citet{Price:2009dc} show that fragmentation in a core with initial mass of 50 M$_sun$ and initial radius of 0.375 pc --possibly not that much different from a low mass star cluster forming core like B59-- proceeds within 1-2 $t_{ff}$ independently of the amount of magnetic and radiative feedback support added to their models.

\par Our analysis confirms the hypothesis that B59-09ab does not show significant fragmentation at the present time. We speculate that the clump is likely on its way to collapse, but it will not form multiple sources to increase the population of the small stellar cluster. The efficiency of formation in the cluster B59 after the collapse of the B59-09 clump will increase only modestly. B59 probably will remain as a small, low mass star cluster with too few stars to survive disintegration by evaporation \citep{Lada:2003aa}. 



\acknowledgments

We want to thank an anonymous referee for a critical reading and a list of useful comments that greatly improved the content of our original manuscript. This study is based on observations carried out with the IRAM 30-m telescope.
IRAM is supported by INSU/CNRS (France), MPG (Germany), and IGN (Spain).
This study makes use of data obtained with instruments from the European
Southern Observatory facilities at La Silla and Paranal. The National Radio Astronomy Observatory is a facility of the National Science Foundation operated under cooperative agreement by Associated Universities, Inc. CRZ acknowledges support from Instituto de Astronom\'ia, UNAM and a repatriation grant from CONACYT, M\'exico. PF and JMG are supported by MICINN grant AYA2008-06189-C03 (Spain) and by AGAUR grant 2009SGR1172 (Catalonia).



{\it Facilities:} \facility{IRAM:30m (MAMBO)} \facility{CAHA:3.5m (OMEGA 2000)} \facility{GBT}



\appendix





\begin{thebibliography}{18}
\expandafter\ifx\csname natexlab\endcsname\relax\def\natexlab#1{#1}\fi

\bibitem[{{Adams} \& {Myers}(2001)}]{Adams:2001aa}
{Adams}, F.~C. \& {Myers}, P.~C. 2001, \apj, 553, 744

\bibitem[{{Alves} {et~al.}(2008){Alves}, {Franco}, \&
  {Girart}}]{Alves:2008bk}
{Alves}, F.~O., {Franco}, G.~A.~P., \& {Girart}, J.~M. 2008{\natexlab{a}},
  \aap, 486, L13

\bibitem[{{Alves} {et~al.}(2007){Alves}, {Lombardi}, \& {Lada}}]{Alves:2007aa}
{Alves}, J., {Lombardi}, M., \& {Lada}, C.~J. 2007, \aap, 462, L17

\bibitem[{{Alves} {et~al.}(2008){Alves}, {Lombardi}, \&
  {Lada}}]{Alves:2008kx}
---. 2008, in {\em Handbook of Star Formation Vol. II}, {ed. Reipurth, B.}, 415

\bibitem[{{Boss}(2009)}]{Boss:2009vn} Boss, A.~P., 2009, \apj, 697, 1940

\bibitem[{{Brooke} {et~al.}(2007){Brooke}, {Huard}, {Bourke}, {Boogert},
  {Allen}, {Blake}, {Evans}, {Harvey}, {Koerner}, {Mundy}, {Myers}, {Padgett},
  {Sargent}, {Stapelfeldt}, {van Dishoeck}, {Chapman}, {Cieza}, {Dunham},
  {Lai}, {Porras}, {Spiesman}, {Teuben}, {Young}, {Wahhaj}, \&
  {Lee}}]{Brooke:2007aa}
{Brooke}, T.~Y., {Huard}, T.~L., {Bourke}, T.~L., et al. 2007, \apj, 655, 364

\bibitem[{{Covey} {et~al.}(2010){Covey}, {Lada}, {Rom{\'a}n-Z{\'u}{\~n}iga},
  {Muench}, {Forbrich}, \& {Ascenso}}]{Covey:2010uq}
{Covey}, K.~R., {Lada}, C.~J., {Rom{\'a}n-Z{\'u}{\~n}iga}, C., {Muench}, A.~A.,
  {Forbrich}, J., \& {Ascenso}, J. 2010, \apj, 722, 971

\bibitem[{{Crutcher}(1999)}]{Crutcher:1999aa}
{Crutcher}, R.~M. 1999, \apj, 520, 706

\bibitem[Crutcher et al.(2004)]{Crutcher:2004aa} Crutcher, R.~M., 
Nutter, D.~J., Ward-Thompson, D., \& Kirk, J.~M.\ 2004, \apj, 600, 279

\bibitem[{{Dapp} \& {Basu}(2009)}]{dapp:2009aa}
{Dapp}, W.~B. \& {Basu}, S. 2009, \mnras, 395, 1092

\bibitem[{{Di Francesco} {et~al.}(2010){Di Francesco}, {Sadavoy}, {Motte},
  {Schneider}, {Hennemann}, {Bontemps}, {Csengeri}, {Balog}, {Zavagno},
  {Andre}, {Saraceno}, {Griffin}, {Men'shchikov}, {Abergel}, {Baluteau},
  {Bernard}, {Cox}, {Deharveng}, {Didelon}, {di Giorgio}, {Hargrave}, {Huang},
  {Kirk}, {Leeks}, {Li}, {Marston}, {Martin}, {Minier}, {Molinari}, {Olofsson},
  {Persi}, {Pezzuto}, {Russeil}, {Sauvage}, {Sibthorpe}, {Spinoglio}, {Testi},
  {Teyssier}, {Vavrek}, {Ward-Thompson}, {White}, {Wilson}, \&
  {Woodcraft}}]{Di-Francesco:2010aa}
{Di Francesco}, J., {Sadavoy}, S., {Motte}, F., et al. 2010, \aap, 518, L91

\bibitem[Fiedler \& Mouschovias(1993)]{Fiedler:1993fk} Fiedler, R.~A., \& Mouschovias, T.~C.\ 1993, \apj, 415, 680 

\bibitem[{{Forbrich} {et~al.}(2009){Forbrich}, {Lada}, {Muench}, {Alves}, \&
  {Lombardi}}]{Forbrich:2009ab}
{Forbrich}, J., {Lada}, C.~J., {Muench}, A.~A., {Alves}, J., \& {Lombardi}, M.
  2009, \apj, 704, 292

\bibitem[{{Franco} {et~al.}(2010){Franco}, {Alves}, \&
  {Girart}}]{Franco:2010ly}
{Franco}, G.~A.~P., {Alves}, F.~O., \& {Girart}, J.~M. 2010, \apj, 723, 146

\bibitem[{{Frau} {et~al.}(2010){Frau}, {Girart}, {Beltr{\'a}n}, {Morata},
  {Masqu{\'e}}, {Busquet}, {Alves}, {S{\'a}nchez-Monge}, {Estalella}, \&
  {Franco}}]{Frau:2010fk}
{Frau}, P., {Girart}, J.~M., {Beltr{\'a}n}, M.~T., {Morata}, O., {Masqu{\'e}},
  J.~M., {Busquet}, G., {Alves}, F.~O., {S{\'a}nchez-Monge}, {\'A}.,
  {Estalella}, R., \& {Franco}, G.~A.~P. 2010, \apj, 723, 1665

\bibitem[Frau et al.(2011a)]{frau11a} Frau, P., Girart, J. M., \& Beltr\'an, M. T. 2011, \aap, in press (arXiv:1112.5319)

\bibitem[Frau et al.(2011b)]{frau11b} ---. 2011, submitted to \aap

\bibitem[Galv{\'a}n-Madrid et al.(2007)]{GalvanMadrid:2007vy} 
Galv{\'a}n-Madrid, R., V{\'a}zquez-Semadeni, E., Kim, J., 
\& Ballesteros-Paredes, J.\ 2007, \apj, 670, 480 

\bibitem[{Krumholz {et~al.}(2007)Krumholz, Klein, \& Mckee}]{Krumholz:2007hs}
Krumholz, M.~R., Klein, R.~I., \& Mckee, C.~F. 2007, \apj,
  656, 959

\bibitem[Lada \& Lada(2003)]{Lada:2003aa} Lada, C.~J., \& Lada, E.~A.\ 2003, \araa, 41, 57

\bibitem[{{Lada} {et~al.}(2008){Lada}, {Muench}, {Rathborne}, {Alves}, \&
  {Lombardi}}]{Lada:2008aa}
{Lada}, C.~J., {Muench}, A.~A., {Rathborne}, J., {Alves}, J.~F., \& {Lombardi},
  M. 2008, \apj, 672, 410

\bibitem[{Longmore {et~al.}(2010)Longmore, Pillai, Keto, Zhang, \&
  Qiu}]{Longmore:2010ei}
Longmore, S.~N., Pillai, T., Keto, E., Zhang, Q., \& Qiu, K. 2010, \apj, 726, 97

\bibitem[{{Lombardi} \& {Alves}(2001)}]{Lombardi:2001aa}
{Lombardi}, M. \& {Alves}, J. 2001, \aap, 377, 1023

\bibitem[{{Mouschovias}(1991)}]{Mouschovias:1991aa}
{Mouschovias}, T.~C. 1991, \apj, 373, 169

\bibitem[Nakamura \& Li(2011)]{Nakamura:2011hs}
F.~{Nakamura} and Z.-Y. {Li}. 2011, in {\em Computational Star Formation},
{ed. J.~Alves, B.~G.~Elmegreen, J.~M.~Girart, \& V.~Trimble}, IAU Symp., 270, 115

\bibitem[{{Offner} {et~al.}(2010){Offner}, {Kratter}, {Matzner}, {Krumholz}, \&
  {Klein}}]{Offner:2010ks}
Offner, S. S.~R., Kratter, K.~M., Matzner, C.~D., Krumholz, M.~R., \& Klein,
  R.~I. 2010, \apj , 725, 1485

\bibitem[{{Onishi} {et~al.}(1999){Onishi}, {Kawamura}, {Abe}, {Yamaguchi},
  {Saito}, {Moriguchi}, {Mizuno}, {Ogawa}, \& {Fukui}}]{Onishi:1999aa}
{Onishi}, T., {Kawamura}, A., {Abe}, R., {Yamaguchi}, N., {Saito}, H.,
  {Moriguchi}, Y., {Mizuno}, A., {Ogawa}, H., \& {Fukui}, Y. 1999, \pasj, 51,
  871

\bibitem[Ossenkopf \& Henning(1994)]{ossen94} Ossenkopf, V., \& Henning, T.\ 1994, \aap, 291, 943 

\bibitem[{Price \& Bate(2007)}]{Price:2007el}
Price, D.~J. \& Bate, M.~R. 2007, \mnras, 377, 77

\bibitem[{Price \& Bate(2009)}]{Price:2009dc}
Price, D.~J. \& Bate, M.~R. 2009, \mnras, 398, 33

\bibitem[{{Rathborne} {et~al.}(2009){Rathborne}, {Lada}, {Muench}, {Alves},
  {Kainulainen}, \& {Lombardi}}]{Rathborne:2009aa}
{Rathborne}, J.~M., {Lada}, C.~J., {Muench}, A.~A., {Alves}, J.~F.,
  {Kainulainen}, J., \& {Lombardi}, M. 2009, \apj, 699, 742

\bibitem[{{Riaz} {et~al.}(2009){Riaz}, {Mart{\'\i}n}, {Bouy}, \& {Tata}}]{Riaz:2009dy}
Riaz, B., Mart{\'\i}n, E.~L., Bouy, H., \& Tata, R. 2009, \apj, 700, 1541


\bibitem[{{Rom{\'a}n-Z{\'u}{\~n}iga} {et~al.}(2009){Rom{\'a}n-Z{\'u}{\~n}iga},
  {Lada}, \& {Alves}}]{Roman-Zuniga:2009aa}
{Rom{\'a}n-Z{\'u}{\~n}iga}, C.~G., {Lada}, C.~J., \& {Alves}, J.~F. 2009, \apj,
  704, 183


\bibitem[{{Rom{\'a}n-Z{\'u}{\~n}iga} {et~al.}(2010){Rom{\'a}n-Z{\'u}{\~n}iga},
  {Alves}, {Lada}, \& {Lombardi}}]{Roman-Zuniga:2010vn}
{Rom{\'a}n-Z{\'u}{\~n}iga}, C.~G., {Alves}, J.~F., {Lada}, C.~J., \&
  {Lombardi}, M. 2010, \apj, 725, 2232

\bibitem[{{Schnee} {et~al.}(2010){Schnee}, {Enoch}, {Johnstone}, {Culverhouse},
  {Leitch}, {Marrone}, \& {Sargent}}]{Schnee:2010uq}
{Schnee}, S., {Enoch}, M., {Johnstone}, D., {Culverhouse}, T., {Leitch}, E.,
  {Marrone}, D.~P., \& {Sargent}, A. 2010, \apj, 718, 306


\bibitem[{{Williams} {et~al.}(1995){Williams}, {Blitz}, \&
  {Stark}}]{Williams:1995aa}
{Williams}, J.~P., {Blitz}, L., \& {Stark}, A.~A. 1995, \apj, 451, 252

\end{thebibliography}

\onecolumn
\clearpage

\clearpage

\begin{deluxetable}{ccccccc}
\rotate
\tablecolumns{7}
\tablewidth{0pc}
\tablecaption{Properties of YSOs counterparts \label{tab:ysos}} 
\tablehead{

\colhead{} &
\colhead{} &
\colhead{} &
\colhead{} &
\colhead{} &
\multicolumn{2}{c}{Gaussian fit}\\
\cline{6-7}\\

\colhead{ID\tablenotemark{a}} &
\colhead{Class\tablenotemark{a}} &
\colhead{Spectral Type\tablenotemark{b}} &
\colhead{Peak Flux} &
\colhead{Mass\tablenotemark{b}}&
\colhead{$\Delta\alpha$,$\Delta\delta$\tablenotemark{c}}&
\colhead{deconvolved size\tablenotemark{d}}\\
\colhead{} &
\colhead{} &
\colhead{} &
\colhead{[mJy/beam]} &
\colhead{[M$_\odot$]} &
\colhead{($\arcsec$)} &
\colhead{($\arcsec$),($\deg$)}
}
\startdata
BHB07-11 & I   & -- & 270.0  & -- & -11.0,\ 55.5 & 18.0$\times$17.0,\ -68.0 \\
BHB07-10 & 0/I & -- & 65.0   & -- & \ 65.0,\ -24.0 & 16.0$\times$16.0,\ 0.0 \\
BHB07-09 & Flat & K5 & 115.0 & 0.77--0.79 & -29.4,\ -133.0 & 13.7$\times$12.1,\ -89.5 \\
BHB07-06 & II & M2 & 15.0 & 0.24--0.62 & \ 15.0,\ -80.0 & 14.0$\times$12.0,\ 79.4 \\
BHB07-07 & Flat& K5 & 15.0 & 0.75--1.16 & \ 15.0,\ -80.0   & 14.0$\times$12.0,\ 79.4 \\

\enddata
\tablenotetext{a}{from \cite{Brooke:2007aa}} 
\tablenotetext{b}{from \cite{Covey:2010uq}} 
\tablenotetext{c}{offsets from center of map at $(\alpha,\delta)$=(17:11:24.0, -27:25:30.0)}
\tablenotetext{d}{indicates major and minor axis, and position angle}
\end{deluxetable}

\begin{deluxetable}{lr}
\tablecolumns{2}
\tablewidth{0pc}
\tablecaption{Barnard 59. Dust emission map. \label{tab:clump}} 
\tablehead{
\colhead{Parameter} &
\colhead{Value}\\
}
\startdata

T$_{k}$\tablenotemark{a} & 10.25 K\\
rms & 4 mJy/beam\\
total flux & 8.28 Jy\\
peak flux & 90.7 mJy/beam\\
Diameter\tablenotemark{b} & 0.11 pc\\
N$_{H_2}$\tablenotemark{c} & 2.96$\times\mathrm{10}^{22}$ cm$^{-2}$\\
n$_{H_2}$\tablenotemark{c} & 1.30$\times\mathrm{10}^{5}$ cm$^{-3}$\\
Mass & 9.19 $M_\odot$\\

\enddata
\tablenotetext{a}{Corresponds to our best fit to the $A_V$ profile, 
not to a measured value}
\tablenotetext{b}{Size of region with emission above $I_\nu$$>$$0.2~I_{\rm \nu, max}$} 
\tablenotetext{c}{Average value over region with emission above $I_\nu$$>$$0.2~I_{\rm \nu, max}$} 
\end{deluxetable}



\end{document}